\documentclass{aa}  
\usepackage{graphicx}
\usepackage{txfonts}
\def\simless{\mathbin{\lower 3pt\hbox
     {$\rlap{\raise 5pt\hbox{$\char'074$}}\mathchar"7218$}}} 
\def\simgreat{\mathbin{\lower 3pt\hbox
     {$\rlap{\raise 5pt\hbox{$\char'076$}}\mathchar"7218$}}} 

\begin{document}
%
   \title{VISIR\,/\,VLT mid-infrared imaging of Seyfert
     nuclei:\,\thanks{Based on VISIR science verification
       observations under ESO/VLT programme 60.A-9244(A).}
     }
     \subtitle{Nuclear dust emission and the Seyfert-2 dichotomy}
   \author{ Martin Haas
          \inst{1}
          \and
          Ralf Siebenmorgen
          \inst{2}
          \and
          Eric Pantin
          \inst{3}
	  \and
          Hannes Horst
          \inst{4,5,6}
          \and
          Alain Smette
          \inst{6}
          \and \\
           Hans-Ulrich K\"aufl
          \inst{2}
          \and
          Pierre-Olivier Lagage 
          \inst{3}
          \and
          Rolf Chini
          \inst{1}
	            }

   \offprints{Martin Haas, haas@astro.rub.de}

   \institute{Astronomisches Institut, Ruhr-Universit\"at Bochum,
              Universit\"atsstra{\ss}e 150\,/\,NA7, 44801 Bochum, Germany
             \and
	     European Southern Observatory, Karl-Schwarzschildstr. 2, 
             85748 Garching b. M\"unchen, Germany 
             \and 
	     DSM/DAPNIA/Service
             d'Astrophysique, CEA/Saclay, 91191 Gif-sur-Yvette, France
             \and
	     Institut f\"ur Theoretische Physik und Astrophysik,
	     Universit\"at zu Kiel, Leibnizstr. 15, 24098 Kiel, Germany
             \and
	     Zentrum f\"ur Astronomie Heidelberg, Institut f\"ur Theoretische
	     Astrophysik, Albert-\"Uberle-Str. 2, 69120 Heidelberg, Germany
             \and
             European Southern Observatory, Casilla 19001, Santiago 19,
             Chile
   }

   \date{Received 21. March 2007; accepted 17. July 2007}

 
  \abstract
  {} 
   { Half of the Seyfert-2 galaxies
     escaped detection of broad lines in their
     polarised spectra observed so far. Some authors have suspected
     that these non-HBLRs contain real Sy2 nuclei without 
     intrinsic broad line region hidden behind a dust torus. 
     If this were true, then their nuclear structure would fundamentally
     differ from that of 
     Sy2s with polarised broad lines: in particular,
     they would not be explained by orientation-based AGN unification. 
     Further arguments for two physically different 
     Sy2 populations have been derived 
     from the warm and cool IRAS F25/F60 ratios. These ratios,
     however, refer to the entire host galaxies and
     are unsuitable to conclusively establish the absence of a nuclear
     dust torus. 
     Instead, a study of the Seyfert-2 dichotomy should be
     performed on the basis of nuclear properties only.
     Here we present the first comparison between
     [OIII]$_{\lambda 5007 \AA}$  and mid-infrared imaging at matching 
     spatial resolution.
     The aim is to check whether the nuclear dust emission
     scales with AGN luminosity as traced by [OIII]. 
   } 
   {During the scientific verification phase of the VISIR 
     instrument at the ESO Very Large Telescope we observed 
     16 Sy1 and Sy2 nuclei at 11.25\,$\mu$m with 0$\farcs$35 spatial 
     resolution (FWHM). 
     We supplement our observations with 
     high-resolution 10--12\,$\mu$m literature data of 58 Seyfert
     galaxies, for 
     which spectroscopic or spectropolarimetric information 
     and far-infrared data are available. 
   }
   { Twelve of the 15 detected sources are unresolved and three sources
     show a dominant unresolved core surrounded by some faint
     knots in an area smaller than 1--2$\arcsec$ radius.
     Our VISIR photometry agrees to better than 15\% with published
     data obtained at 1$\farcs$5--5$\arcsec$ spatial resolution.  
     Exploring the Seyfert-2 dichotomy we find that the distributions of 
     nuclear mid-infrared\,/\,[OIII] luminosity ratios are
     indistinguishable for Sy1s and Sy2s  with and
     without detected polarised broad lines and
     irrespective of having warm or cool IRAS F25/F60 ratios.  
     We find no evidence for the existence of a population of 
     real Sy2s with a deficit of nuclear dust emission.  
     Our results  suggest
     1) that all Seyfert nuclei  possess the same
     physical structure including the putative dust torus and
     2) that the cool IRAS colours are caused by a low contrast
     of AGN to host galaxy.
     Then the Seyfert-2 dichotomy is explained
     in part by unification of non-HBLRs with narrow-line Sy1s
     and to a larger rate by observational biases caused by a low
     AGN/host contrast and/or an unfavourable scattering geometry.  
   } 
  {}

   \keywords{Galaxies: nuclei, active, Seyferts - Infrared: galaxies}

   \maketitle
%

\section{Introduction}
The nuclei of Seyfert galaxies are grouped into Sy1 and Sy2 
depending on the presence of broad emission lines in their optical
spectra (Khachikian \& Weedman 1974).
According to the 
unified model Sy1 and Sy2 nuclei differ only with respect 
to our line of sight. In Sy2s the broad line region (BLR) 
is hidden by a dusty torus-like structure seen edge-on. 
Crucial evidence for the unified model comes
from the detection of polarised broad lines in those Sy2 galaxies, 
where a 
"scattering mirror" off the torus plane allows a
direct view of the region inside the torus 
(Antonucci 1993).
Also, in some cases broad Pa$\beta$ lines were found by  
ordinary infrared spectroscopy penetrating the dust column
(e.g. Veilleux et al. 1997).
The presence of obscuring material is most directly inferred from
large X-ray absorbing columns as well as the thermal emission from the
optical-UV energy intercepted and re-radiated in
the mid-infrared (MIR, 3--40$\mu$m). 
Note that the unified model refers to the structure of an AGN with 
the dusty molecular torus belonging to its basic components, but 
does not make a statement about the host galaxies (Antonucci 2002).

So far about half of 
the Sy2s show evidence for a hidden broad line region (HBLR)
and half of them do not (e.g. compilation by Gu \& Huang 2002, 
Moran 2007).
Such a dichotomy between HBLR and non-HBLR Sy2s
could either (1) result from the existence of real Sy2s
without intrinsic BLR
or (2)
be due to an observational bias.
In order to establish the origin of 
the Seyfert-2 dichotomy, numerous studies 
have been performed, but with controversial results.

Theoretical studies suggests that there are limits to the
existence of a BLR, in particular at low AGN luminosity
(Nicastro 2000, Nicastro et al. 2003).
At the extreme end of the Sy1 population some sources show rather
narrow H$\beta$ lines (FWHM $<$ 2000 km/s), but extraordinarily
strong FeII lines and steep X-ray spectra not found in Sy2s.
For the narrow-line Seyfert-1 galaxies (NLS1s) 
some orientation-based type-2 counterparts have been found 
(e.g. Nagar et al. 2002, Dewangan \& Griffiths 2005, 
Zhang \& Wang 2006). 
Since NLS1s constitute less than
15\% of the optically selected Sy1 population,
some but not all of the non-HBLR Sy2s may be misoriented NLS1s. 

Arguments for an origin of the Seyfert-2 dichotomy as an observational bias are
numerous.
Modelling X-ray spectra can provide column densities N$_{\rm H}$ as 
valuable constraints (Alexander 2001, Gu et al. 2001), but depends
critically on assumptions about the central geometry.
Hence, any inference on low N$_{\rm H}$
may be pretended by X-ray scattering (e.g. Ghosh et al. 2007) and
should be corroborated by other findings. 
There is no doubt that dust lanes 
may obscure not only the nucleus but also the scattering 
mirror necessary for the detection of a HBLR. Hence, 
only under favourable circumstances one may expect detectable
scattered light at all from a hidden AGN (Miller \& Goodrich 1990,
Heisler et al. 1997, Gu et al. 2001). 
The AGN-typical emission line [OIII]$_{\lambda 5007 \AA}$
(henceforth denoted [OIII]) has an average equivalent width which
is higher for HBLRs than for non-HBLRs, indicating a stronger AGN/host
contrast in HBLRs (Lumsden et al. 2001, Moran 2007).
Generally, it is a challenge for spectropolarimetric observations 
to discern the few percent BLR signature of a relatively faint AGN against
a luminous host galaxy (Alexander 2001, Gu et al. 2001,
Lumsden \& Alexander 2001, Lumsden et al. 2001).
Although still half of the most nearby
Sy2s resisted HBLR detection even with
Keck spectropolarimetry, such sensitive observations are
revealing broad lines in sources,
which were previously classified as non-HBLRs using smaller 
telescopes. This reminds us to take care when interpreting 
spectropolarimetric non-detections of broad lines (Moran 2007). 

From his Lick-Palomar spectropolarimetric survey of the
CfA and 12\,$\mu$m Seyfert
samples Tran (2001, 2003) has found that compared with HBLRs the 
non-HBLRs show lower [OIII] luminosity, lower [OIII]/H$_{\beta}$
excitation ratios and cooler IRAS 25\,$\mu$m\,/\,60\,$\mu$m 
colours F25/F60\,$<$\,0.25. 
Most Seyfert galaxies exhibit warm F25/F60,
but some have cool colours.
Guided by the widespread belief that cool F25/F60 indicates 
a lack of adequate nuclear dust emission,
Tran naturally concluded that most if not all non-HBLRs
are real Sy2s,
and not misaligned Sy1s. 
However, the host galaxies have a size of about 1$\arcmin$ 
so that nuclei and hosts are not separated by the IRAS beam. 
Hence, {\em nuclear} optical properties were compared with 
{\em extended} infrared ones possibly dominated by the host
so that the conclusions about the missing nuclear dust torus
in non-HBLRs should be checked using adequate nuclear data. 

If we focus on the BLR and the dust torus as basic components
of the structure of an AGN, then intrinsically an AGN may
belong to one of the four formal cases: 
\begin{itemize}
\item[1)] with BLR and with dust torus
\item[2)] with BLR, but without dust torus
\item[3)] without BLR, but with dust torus
\item[4)] without BLR and without dust torus.
\end{itemize}
The sources of case 1 are Sy2s with HBLR and Sy1s,
while sources of case 2 have not been observed
so far (they would be Sy1s without nuclear dust). 
If non-HBLRs do not belong to
case 1, then they are sources of case 3 and/or 4 and nuclear MIR
observations should be able to distinguish between case 3 and 4.
Therefore, we define a "naked" AGN to be 
free of surrounding dust (in analogy to the terminology of T Tauri
stars). The assessment of whether
nuclei are naked or not requires a suitably chosen reference quantity
and reference sample.
We here combine new nuclear MIR 11.25\,$\mu$m observations at the VLT with
published nuclear 10--12\,$\mu$m photometry of several Seyfert type
samples, and compare them with
[OIII] literature data as reference quantity.
The [OIII] emission arises from the moderately extended narrow-line
region (NLR).
Our test assumes that the [OIII] emission can be regarded with
little reservation as isotropic measure of the intrinsic AGN power so
that it can be used for suitable normalisation. 
The aim here is to check whether there exist naked nuclei
among Seyfert galaxies and in particular among non-HBLRs.
At a first guess we expect that 
the L$_{\rm MIR}$\,/\,L$_{\rm [OIII]}$ ratio of a naked nucleus
lies below the distribution of 
that ratio found in most Sy1s or HBLR Sy2s.
Furthermore if nuclear dust emission is missing in non-HBLR sources
(case 4), this argues against a hidden BLR and in favour of real
Sy2s. 
On the other hand, if the nuclear MIR\,/\,[OIII] ratio has the same
distribution for both cool non-HBLR and warm HBLR Sy2s,
then this suggests that both types are similarly surrounded by a dust torus
(case 1 or 3). Furthermore if 
case 3 may be rejected with the help of other arguments,
then both Sy2 types possess the same physical AGN structure. 

\section{VISIR science verification observations}

\begin{table}
\caption{
  VISIR 11.25\,$\mu$m photometry
  with typical errors $\sim$10\%
  and literature 10\,$\mu$m (N-band) photometry. 
}  
\label{table_visir}      
\begin{center}
\begin{tabular}{lrrc}
object        & VISIR          & literature      & reference$^{*}$    \\
              & mJy            &  mJy\parbox{0cm}{$^{a}$}  &             \\
\hline
  CentaurusA  &  946           & 1000                              & 1        \\
 ESO141-G055  &  169           &  166                              & 2             \\
      IC5063  &  752           &  920                              & 1	       \\
 MCG-3-34-64  &  674           &  594                              & 3               \\
 MCG-6-30-15  &  392           &  383                              & 4	       \\
     Mrk1239  &  660           &  600                              & 5	       \\
      Mrk509  &  235           &  240                              & 5             \\
      Mrk590  &   90           &  100                              & 6		       \\
     NGC2992  &  312           &  339                              & 7                 \\
     NGC3783  &  645           &  689\parbox{0cm}{$^{b}$}          & 7               \\
NGC4507       &  589           &  600\parbox{0cm}{$^{c}$}          & 6		       \\
     NGC5427  & $<$2.3         &   $\sim$10                        & 8	          \\
     NGC5995  &  332           &  300                              & 9             \\
     NGC7213  &  250           &  261                              & 4	       \\
     NGC7314  &   74           &  75, $<$100\parbox{0cm}{$^{c}$}   & 10, 6  \\
  PG2130+099  &  160           &  174, 130                         & 10, 9          \\
\hline
\end{tabular}
\end{center}
$^{*}$  The references are:\\
1 = Siebenmorgen et al. (2004), 
2 = Rieke \& Low (1972),    
3 = Gorjian et al. (2004), 
4 = Glass et al. (1982), 
5 = Maiolino et al. (1995), 
6 = this work using Spitzer IRS, 
7 = Roche et al. (1991), 
8 = this work using Spitzer IRAC \& MIPS, 
9 = Galliano et al. (2005),  
10 = Horst et al. (2006).

$^{a}$ aperture $\sim$5$\arcsec$, 
  except $\sim$1$\farcs$5
  for Siebenmorgen et al. (2004), Gorjian et al. (2004), Galliano et al. (2005),
  and $\sim$0$\farcs$5 for Horst et al. (2006).

$^{b}$ flux at 12\,$\mu$m.

$^{c}$  corrected for flux loss due to slit offset, see text. 

\end{table}

We have observed about two dozen Seyfert
galaxies from several cataloges during the scientific
verification phase of VISIR.
VISIR is the VLT imager
and spectrograph for the mid--infrared (Lagage et al. 2004,
Pantin et al. 2005), mounted on the Cassegrain focus of the
VLT Unit Telescope 3
(Melipal). In order to verify the capabilities of VISIR,
the sources were selected to cover a broad range of
properties. The AGN exhibit faint as well as strong starburst 
contributions and have extended to so far known unresolved nuclei.
The science verification sources do not form a
homogeneous AGN sample.
Here we consider those 16 sources suited to 
adress the Seyfert-2 dichotomy. 

The imaging data were obtained
during Oct. 2004 -- Feb. 2005 
through the PAH2 filter (11.25\,$\pm$\,0.6\,$\mu$m) under good and stable 
weather conditions.
The optical seeing was better than 1$\arcsec$ and the objects were observed 
at airmass $<$\,1.4 (1.1 on average).
To suppress the
background, secondary mirror chopping was
performed in North-South direction with an amplitude of 16$\arcsec$ at
a frequency of 0.25\,Hz.
Nodding was applied every 30\,s using telescope offsets of
16$\arcsec$ in East-West direction. The pixel scale was
0.127 arcsec/pixel resulting in a 32$\farcs$5 field of view.
The detector integration
time was 25\,ms. Total source integration time was 20\,min.
All observations were bracketed by photometric standards
(from http://www.eso.org/instruments/visir/tools/).
The elementary images are coadded in real-time to obtain
chopping-corrected data. Then the
nodding positions are combined to create the final image.
VISIR images may show stripes randomly triggered by some high-gain pixels.
They are removed by a dedicated reduction method (Pantin 2007, in prep.).

\begin{figure}
  \vspace{-3mm}

  \hspace{-6mm}
  \includegraphics[angle=0,width=9.5cm,clip=true]{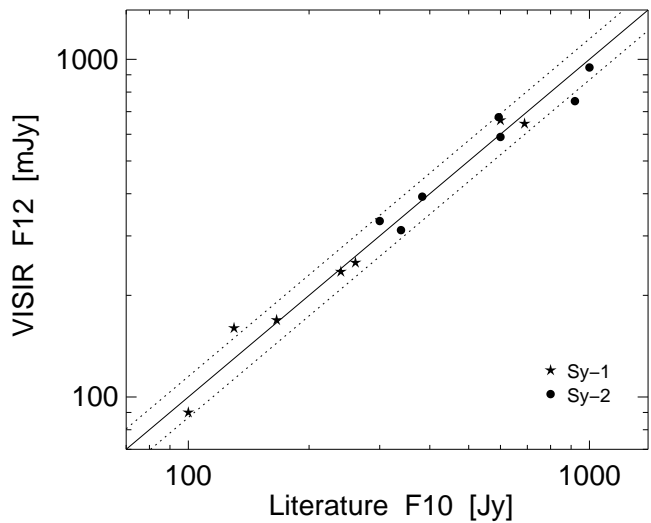}

  \caption{Comparison of VISIR photometry with literature results. The
  dotted lines indicate 15\% deviation from unity. 
    \label{fig_visir}
  }
  \vspace{-3mm}
  
\end{figure}

Twelve of the 15 detected sources are unresolved
(FWHM\,=\,0$\farcs$35), and three sources
show a dominant unresolved core surrounded by some
faint knots in an area smaller than 1--2$\arcsec$ radius.
We find that the VISIR 11.25\,$\mu$m photometry is consistent 
with published measurements (Table\,\ref{table_visir}).
In four cases we derived 11-12\,$\mu$m
photometry from archival Spitzer IRAC/MIPS/IRS data at
$\sim$5$\arcsec$ resolution.
Notably, in two cases (marked in Table\,\ref{table_visir}) the
source had an offset of $\sim$1$\arcsec$ to the IRS slit center
(slit width $\sim$4$\arcsec$),
leading to about 30\% flux loss we corrected for. 
Figure\,\ref{fig_visir} illustrates the excellent agreement
of our 
VISIR photometry with other observations.
The similarity of the VISIR fluxes with those
measured using apertures of 1$\farcs$5 -- 5$\arcsec$ suggests that
essentially the entire 10--12\,$\mu$m flux of our sources
arises from very compact nuclear areas (FWHM$\sim$0$\farcs$35).

One caveat has to be mentioned, when chopping
in a structure-rich emission. 
One object (NGC\,5427) has not been detected by VISIR with a formal 
3-$\sigma$ upper limit of only 2.3\,mJy for an unresolved point
source. 
This object has not been observed with other ground-based MIR
arrays, but with the Spitzer Space Telescope at
3.6 -- 24 $\mu$m.
Despite the lower spatial resolution (by a factor of about ten),  
the Spitzer images reveal a nuclear point source
(FWHM\,$\sim$\,3$\arcsec$) and extended ring
like emission at about 10$\arcsec$ separation from the 
nucleus.
From the Spitzer SEDs we estimate that the 3$\arcsec$ nuclear 11.25\,$\mu$m
flux of NGC\,5427 should be at least 10\,mJy, four times higher
than the formal 0$\farcs$5 VISIR upper limit of 2.3\,mJy.
This discrepancy disappears, if the nuclear emission is unresolved by
Spitzer and resolved by VISIR but has an insufficient surface
  brightness to be detected.  
In the following comparison with samples of lower resolution,
we adopt for NGC\,5427 the upper limit of 10\,mJy.

\begin{figure}
  \vspace{-3mm}
  
  \hspace{-6mm}
  \includegraphics[angle=0,width=9.5cm,clip=true]{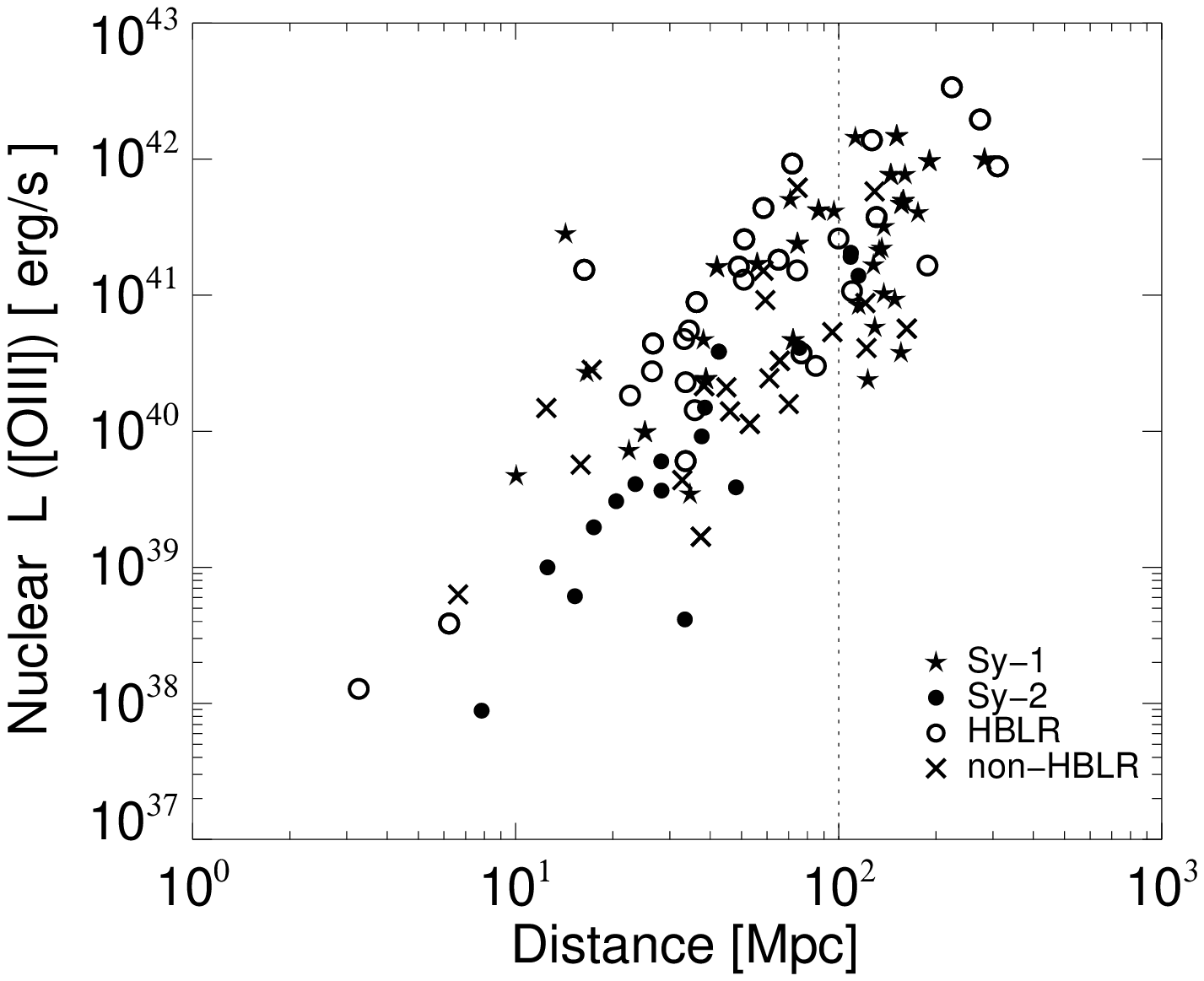}
  \vspace{-8mm}

  \hspace{-6mm}
  \includegraphics[angle=0,width=9.5cm,clip=true]{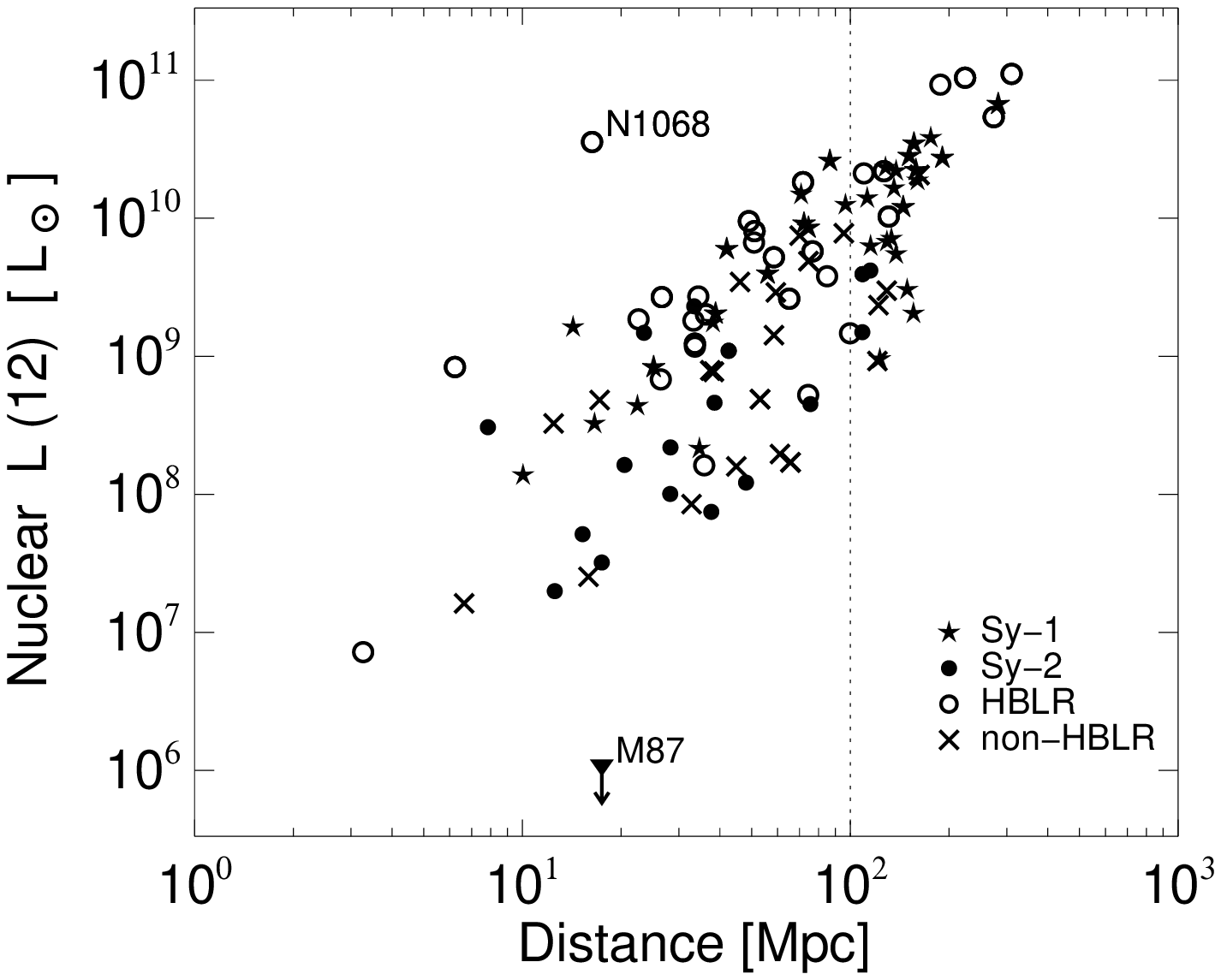}
  \vspace{-8mm}

  \hspace{-6mm}
  \includegraphics[angle=0,width=9.5cm,clip=true]{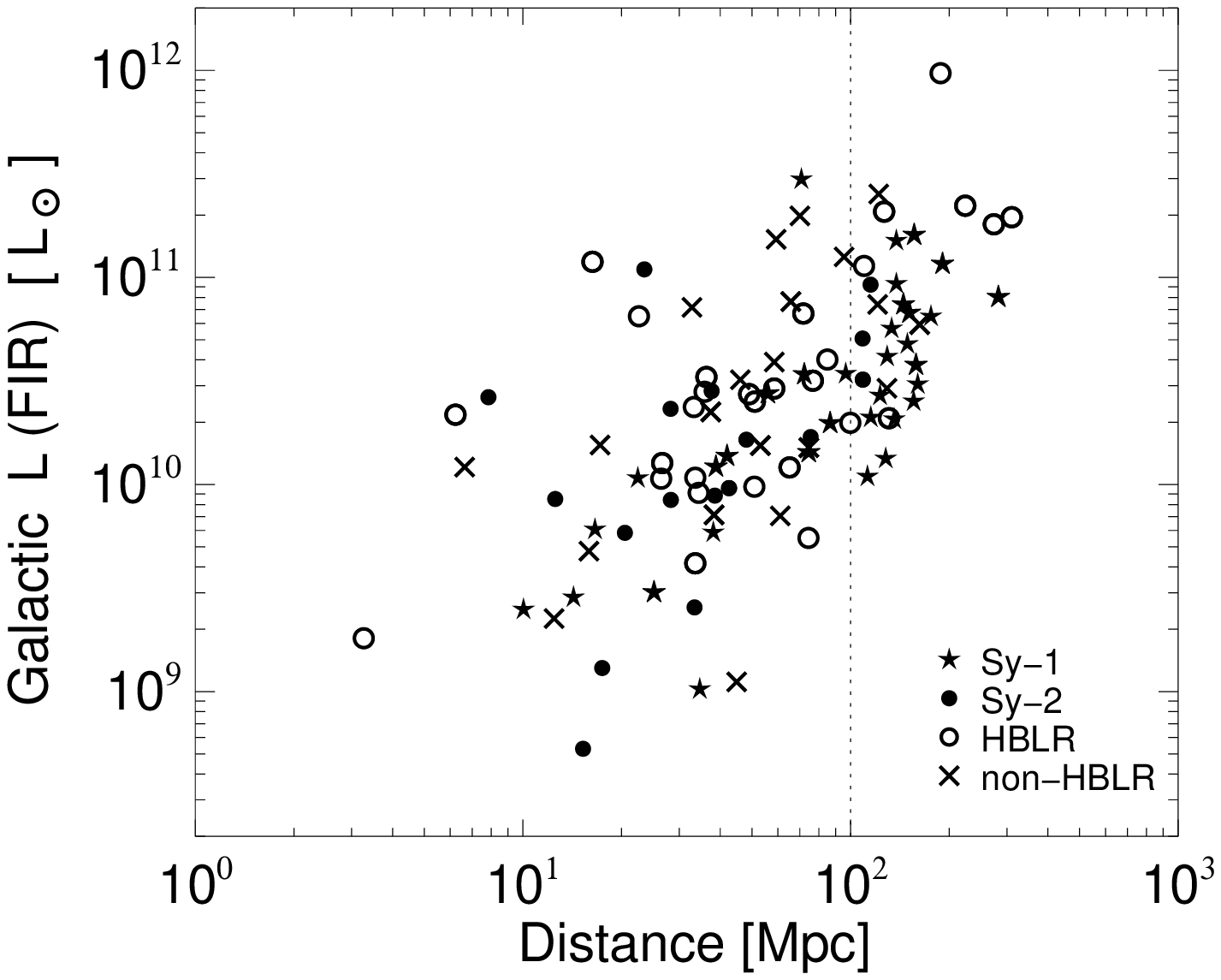}
  \vspace{-8mm}

  \caption{Luminosities versus distance:  
    L [OIII]$_{\lambda 5007}$  (top), 
    L (MIR) (middle), and L (FIR) (bottom).
    The data are shown for the entire sample complied from our
    VISIR observations and the literature.
    We distinguish between Sy1 and Sy2 and additionally mark HBLR/non-HBLR.
    For the analysis of the Sy2 dichotomy we applied a distance cut
    and used only sources at d $\simless$ 100 Mpc (dotted vertical line). 
    \label{fig_luminosities}
  }
  
\end{figure}

\section{Sample selection and data base}

Apart from the science verification issue, our VISIR sources were originally
selected for comparison with X-ray and sub-mm samples
(results in preparation by
Horst et al. and Siebenmorgen et al., respectively) 
so that part of them lacks information needed to study
the Seyfert-2 dichotomy. 
Therefore, we supplement the VISIR sample with literature data. 
So far no data
base with homogeneously observed spectropolarimetry 
and high-resolution MIR photometry exists for a well defined
complete Seyfert sample.

We here selected our sample by starting with
all Sy2 sources having spectropolarimetric
information (Heisler et al. 1997, Lumsden et
al. 2001, 2004, Moran et al. 2000, 2001, 2007, Tran 2003).
They were mostly drawn from the local (cz $<$ 3100 km/s, Ulvestad \&
Wilson 1984), the CfA (Huchra \& Burg 1992) and the 
12\,$\mu$m (Rush et al. 1993) Seyfert samples.
We included all Sy1 galaxies from these samples
as well as the remaining Sy2s without spectropolarimetry, in order to compare 
them with HBLRs and non-HBLRs. 

We cross-correlated this optical data base with 
high-resolution FWHM $\simless$ 1$\arcsec$ array observations at 10--12\,$\mu$m
(Gorjian et al. 2004, 
Siebenmorgen et al. 2004 and Galliano 2005), as well as 
 10\,$\mu$m photometer measurements with $\sim$5$\arcsec$ aperture
(Maiolino et al. 1995 and references therein).
We did not scale the MIR fluxes to a
common wavelength, since such corrections rely on 
assumptions about the spectral slope and would be small either.   
As shown by Gorjian et al. (2004), for 58 of their 62
detected sources virtually the entire flux seen
in the $\sim$5$\arcsec$ aperture arises from the nuclei unresolved to FWHM
$\simless$ 1$\arcsec$, corresponding to about 500 pc resolution for a
source at 100 Mpc distance.  
The MIR observations cannot resolve the dust torus so that
dust clouds in the NLR (e.g. Galliano et al. 2005a) or 
nuclear starbursts may contribute as well, but we can confine the
analysis to the nuclear emission largely free of contamination by the
host galaxy. Dust enshrouded starbursts may contribute to the nuclear MIR
emission, as indicated in some cases resolved with VISIR
(Wold \& Galliano 2006), but 1) 
such sources stand out in the distributions shown below, and 2) the
resolved circumnuclear starbursts are faint relative to the unresolved
nuclei (e.g. Wold \& Galliano 2006). 

We complemented and interpolated missing IRAS 
12-100\,$\mu$m photometry  as far as available 
by ISO and Spitzer photometry. 
We excluded sources without FIR data
(i.e. lying in sky areas not scanned/observed by IRAS/ISO/Spitzer),
sources with LINER spectra, and double
nuclei with unknown location of the MIR data. 
This results in 34 Sy1s (types 1.0--1.5) as well as 
66  Sy2s (types 1.8--2.0), 29 with and 20 
without detected hidden broad lines and 17 without
spectropolarimetry. Table\,\ref{table_sample} lists the sample
parameters and Table\,\ref{table_sources} the source parameters.
Figure\,\ref{fig_luminosities} shows for each
type the distribution of the [OIII],
12\,$\mu$m and FIR (60 and 100\,$\mu$m) luminosity versus distance.
The implications from the luminosities are discussed further below 
(Sect.\,\ref{section_seds} and \ref{section_extinction}).
Since the VISIR data points lie randomly distributed
across the entire sample, they are not marked with extra symbols, 
in order to keep the number of plot symbols manageable. 

\begin{table*}
\caption{Sample parameters. Number of sources, as well as logarithmically averaged values discussed.
  3-$\sigma$ upper limits were treated as detections. 
}  
\label{table_sample}      
\begin{center}
\begin{tabular}{@{\hspace{0.5mm}}l@{\hspace{2.5mm}}|c@{\hspace{2.5mm}}|@{\hspace{2.5mm}}c@{\hspace{2.5mm}}|@{\hspace{2.5mm}}c@{\hspace{2.5mm}}|@{\hspace{2.5mm}}c@{\hspace{2.5mm}}r}
(1)                         &(2)      & (3)    & (4)  & (5)       \\
                            & Sy1    & Sy2$^{\rm a}$  & HBLR    & non-HBLR \\
\hline
Total number of sources            & 34 ~~~ (4)$^{\rm n}$    & 17   &  29     &  20      \\
thereof used                 & 17 ~~~ (2)$^{\rm n}$    & 12   &  21     &  16      \\
\hline
                            &  1.quart, median, 3.quart       & 1.quart, median, 3.quart         &  1.quart, median, 3.quart     &    1.quart, median, 3.quart       \\
                            &   mean~~~~~~dex      &   mean~~~~~~dex        &   mean~~~~~~dex     &   mean~~~~~~dex         \\
                            &         &      &         &          \\
\hline
L([OIII])  [log (erg/s)]                  & 40.38~~~41.21~~~41.45 & 39.29~~~39.96~~~40.58  & 40.57~~~40.95~~~41.41 &  40.05~~~40.33~~~40.96  \\
& 40.76~~~~~~0.72       &  39.73~~~~~~0.65       & 40.78~~~~~~0.85     &   40.37~~~~~~0.68   \\
Nuclear L(12$\mu$m) [log L$_\odot$]        & 8.92  ~~~9.60  ~~~9.96      & 7.71  ~~~8.65  ~~~9.04        &  9.26  ~~~9.43  ~~~9.91    &  8.23  ~~~8.90  ~~~9.54       \\
                             & 9.34  ~~~~~~0.71        &  8.22  ~~~~~~0.68       &  9.41  ~~~~~~0.77    &  8.81  ~~~~~~0.74   \\
L(FIR)$^{\rm b}$ ~~~ [log L$_\odot$]           & 9.48 ~~~10.14 ~~~10.44        &   9.93 ~~~10.22 ~~~10.45      & 10.03 ~~~10.37 ~~~10.52     &    9.85 ~~~10.51 ~~~10.88       \\
                            &   9.93  ~~~~~~0.49      &  9.96  ~~~~~~0.69       &  10.25  ~~~~~~0.43    &  10.29  ~~~~~~0.66   \\
                            &         &         &      &          \\
\hline
Nuclear   F12/[OIII]       & 11.82 ~~~ 11.93  ~~~12.28        &  11.57 ~~~ 11.85 ~~~ 12.28    & 11.75  ~~~12.07  ~~~12.46        &  11.26  ~~~11.86  ~~~12.52        \\
  ~~~~~~  log (Jy/erg/s/cm$^{\rm 2}$)                          & 11.94   ~~~~~~0.39               &  11.84   ~~~~~~0.50    &  11.99  ~~~~~~ 0.43              &  11.79   ~~~~~~0.65          \\
                            &         &      &         &          \\
\hline
Nucl./Gal. F12$^{\rm c}$    &         &  0.06  ~~~    0.13   ~~~   0.15    &  0.42   ~~~   0.50  ~~~    0.60       &  0.04   ~~~   0.25  ~~~    0.72        \\
                            & 0.32  ~~~~~~0.03      &  0.08 ~~~~~~0.44   &    0.29 ~~~~~~0.55    &  0.13 ~~~~~~0.73       \\
                            &         &      &         &          \\
\hline
\end{tabular}
\end{center}
$^{\rm n}$  number of NLSy1 are listed in brackets 

$^{\rm a}$  without spectropolarimetry

$^{\rm b}$ = L(60$\mu$m \& 100$\mu$m)

$^{\rm c}$ only cool sources with F25/F60 $<$ 0.25 
\end{table*}



Since the actual dichotomy, as to whether non-HBLRs are real Sy2s or not,
tends to fall at rather modest AGN luminosity we excluded the most
luminous sources from our analysis by applying a distance
cut at d = 100 Mpc.
This results in 15 broad-line Sy1s, 2 narrow-line Sy1s,
22 HBLRs, 16 non-HBLRs and 14 Sy2s without
spectropolarimetry (Table\,\ref{table_sample}).
They cover 23/25 of the local Ulvestad \& Wilson
sources (two Sy2s were excluded because of missing FIR data) and 42/42 of the
nearby CfA Seyferts.
The sample contains also some narrow-line Sy1s as marked in
Table\,\ref{table_sources}; because they do not
differ from broad-line Sy1s in any properties analysed here,
we do not plot them with extra symbols in the diagrams. 
Our sample is not homogeneously observed, 
but it can be considered as fairly random selection suited to study
the Sy2 dichotomy.

 \section{Results and Discussion}

 \subsection{Nuclear dust emission}
 \label{section_nuclear_dust_emission}
   
\begin{figure}
  \vspace{-3mm}
  
  \hspace{-6mm}
  \includegraphics[angle=0,width=9.5cm,clip=true]{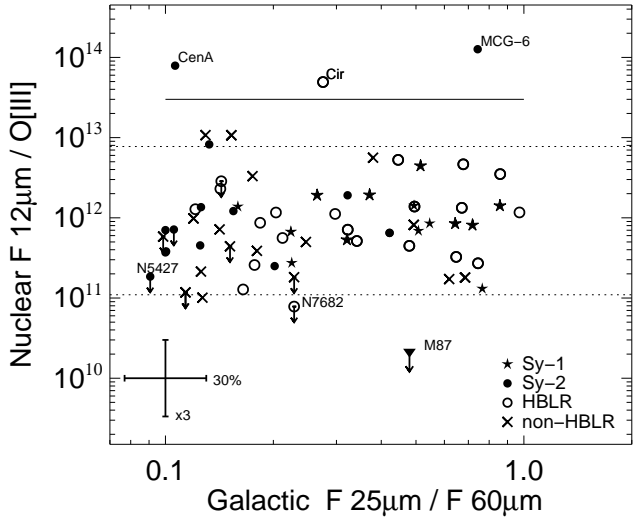}

  \caption{Nuclear MIR flux normalised by [OIII] versus IRAS
    25$\mu$m/60$\mu$m colours from the entire galaxies.
    F12 / [OIII] is in units  Jy / 10$^{\rm -16}$  erg/s/cm$^{\rm 2}$,
    as listed in Table\,\ref{table_sources}. 
    Symbols with arrows denote 3-$\sigma$
    upper limits. The dotted  horizontal lines mark the 3-$\sigma$ range
    around the mean flux ratio for HBLRs. 
    The solid horizontal line (at y = 3\,$\cdot$\,10$^{\rm 13}$)
    markes the transition to sources with the highest
    MIR/[OIII] ratio.
    They are known
    to be contaminated by strong dust enshrouded starbursts and excluded
    from the  analysis.
    \label{fig_f12_to_oiii_vs_iras}
  }
\end{figure}

Figure\,\ref{fig_f12_to_oiii_vs_iras} shows the nuclear MIR flux
normalised by [OIII]
plotted against the IRAS 25$\mu$m/60$\mu$m colours.
All along the range of IRAS colours the bulk of Sy sources lies in the
same MIR/[OIII] range. This is also the case for the different Sy
types,.  
We note that the few narrow-line Sy1s of our sample
fall in the same range covered by Sy1s. 
Three sources (Cen\,A, Circinus and MCG-6-30-15)
show exceptionally high MIR\,/\,[OIII] ratios.
They are known to be contaminated by strong dust
enshrouded starbursts 
and we have excluded them from the analysis.
This decision does not affect the conclusions of this paper. 

The statistics for the MIR\,/\,[OIII] distributions
are listed in Table\,\ref{table_sample}.  
Taking into account the broad dispersions there is no statistically
significant evidence that any of the Seyfert types
shows different MIR\,/\,[OIII] distributions.
In Fig.\,\ref{fig_f12_to_oiii_vs_iras} the dotted  horizontal lines
illustrate, for example, the 3-$\sigma$ range
around the mean flux ratio for HBLRs. 
Only one source (NGC\,7682) lies slightly below the 3-$\sigma$ range.
Remarkably it is a HBLR indicating that the unified model is compatible with 
rather low MIR\,/\,[OIII] values and that naked AGN may have to be
searched for at MIR\,/\,[OIII] $<<$ 10$^{\rm 11}$ Jy / erg/s/cm$^{\rm 2}$. 
NGC\,5427, for which we used the 12\,$\mu$m upper flux limit of 10\,mJy
inferred from Spitzer, shows relatively little nuclear dust emission. But
even when taking the VISIR upper limit of 2.3\,mJy the 
evidence that NGC\,5427 is a naked Sy2 is still marginal. To our
knowledge this source
has not yet been observed by spectropolarimetry. 

For comparison and to get an impression, where a naked Sy has to be
looked at,   
also the location of M\,87 is plotted. Its radio jet
dominates the nuclear MIR flux (Fig.\,2 
in Whysong and Antonucci 2004); we here conservatively adopt
that at most 5\% of the entire nuclear 11.7\,$\mu$m flux  of 13\,mJy
is due to dust.
Then M\,87 lies about 15-$\sigma$  below the average MIR/[OIII] ratio of each 
Seyfert type.
Note already M\,87's exceptionaly low nuclear L$_{\rm 12 \mu m}$ in 
Figure\,\ref{fig_luminosities}, middle. 
Using MIR data from Subaru and Spitzer, Perlman et al. (2007) reach
the same conclusions about M\,87 as Whysong and Antonucci (2004). 
Also, M\,87 has rather a LINER than a high-excitation [OIII] bright Seyfert
spectrum. If it were a Seyfert with a stronger [OIII], we expect
it to be placed at even lower MIR/[OIII]. 
In addition to Whysong and Antonucci's diagnosis, which involved
ADAF and jet models, our comparison with Seyfert galaxies
argues strongly in favor of the naked AGN nature of M\,87. 
And looking the other way around, we conclude that  
none individual of the Seyfert galaxies falls sufficiently below the
low end of the MIR\,/\,[OIII] distributions requird to provide
evidence for being a naked AGN.

While the statistical rms of the MIR and [OIII] measurements
is in the order of 10-30\%, we here adopted 
a factor 3 as uncertainty for MIR/[OIII] in order to account also for
systematic effects. For example 
we used the [OIII] fluxes as observed and neither extinction nor
aperture corrected, since such corrections are uncertain and depend on
assumptions about the source geometry.
Nevertheless we have performed several tests  
to correct for extinction in the [OIII] fluxes on the basis
of the H$\alpha$\,/\,H$\beta$ ratio. With and without extinction correction 
the studied distributions are quite broad and  do not reveal any reliable
trends which would be different
from those already seen in Fig.\,\ref{fig_f12_to_oiii_vs_iras}.
One reason for the apparent "failure" of [OIII] extinction corrections
using H$\alpha$\,/\,H$\beta$ ratios may be that the published fluxes
refer to
observed values.
So far they have not been corrected for potentially significant
stellar absorption, with exception of a few ($\sim$20) sources of our
sample (Gu et al. 2006).
As regards [OIII] flux losses due to small slit widths,    
we did not find any trends of MIR/[OIII] with distance. This
suggests that aperture effects are similar for all sources and
largely cancel out in the flux ratios.  

As an alternative to potentially incomplete or erroneous
extinction and aperture corrections, we discuss the effects any
extinction may have on the MIR/[OIII] ratios and on our conclusions:

1) If a source is substantially obscured at [OIII] and not at MIR wavelengths,
it will be shifted towards
higher MIR\,/\,[OIII] values.
But then by assumption this source cannot be naked, since there must be
nuclear dust to obscure the [OIII] emission.
The dust must be located in a torus/disk like structure, in order to
explain the bipolar morphology of the [OIII] emission observed even in
some non-HBLRs (for example Mrk\,573 or NGC\,1386, see Schmitt et al. 2003).

2) Many Sy2 nuclei, also some among our sample,
show the well known silicate 9.7\,$\mu$m absorption.
If its emission becomes optically thick at MIR wavelengths
with increasing (i.e. more edge-on) inclination of the torus,
one may expect that the Sy2s exhibit a lower
MIR\,/\,[OIII] distribution than the Sy1s. 
But in that case also the nuclear [OIII] emission may be affected by
extinction. This was shown for powerful 3CR radio galaxies
by polarised [OIII] (di Serego Aligieri et al. 1997)
and by suppressed [OIII]/[OIV]$_{\rm 25.9 \mu m}$ (Haas et al. 2005) 
so that the net MIR\,/\,[OIII] ratio tends to higher values.
Future Spitzer MIR spectra will provide further clues to this issue.

These considerations about extinction lead us to conclude 
that the basic results derived 
here from observed MIR\,/\,[OIII]
flux ratios are valid. 
The important conclusions from the MIR\,/\,[OIII] distributions are 1) that
there is no clear indication of naked Seyfert nuclei  
and 2) that the distributions of cool non-HBLRs are comparable to those
of broad- and narrow-line Sy1s and warm HBLRs.


\subsection{Mid- and far-IR contribution of AGN and host}
\label{section_seds}

\begin{figure}
  \vspace{-3mm}

  \hspace{-6mm}
  \includegraphics[angle=0,width=9.5cm,clip=true]{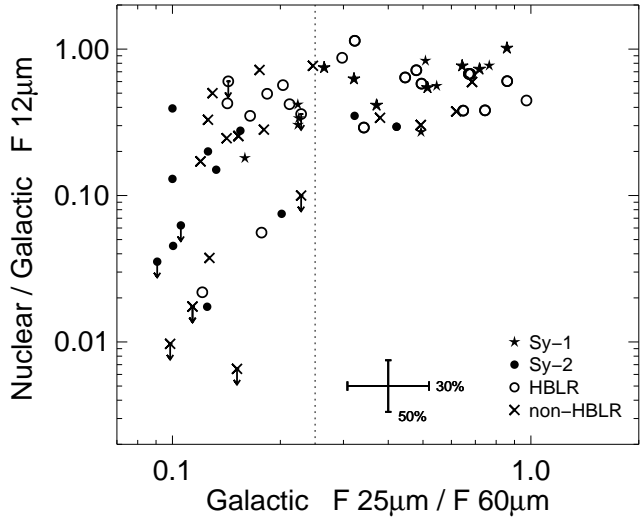}
  \caption{
    Nuclear MIR flux concentration versus IRAS
    25$\mu$m/60$\mu$m colours.
    Symbols with arrows denote 3-$\sigma$ upper limits.
    The vertical dotted line markes the separation between cool and
    warm sources. 
    \label{fig_mir_concentration}
  }
\end{figure}

An important corollary from Figure\,\ref{fig_f12_to_oiii_vs_iras} is:
the nuclear dust emission is independent of the 
IRAS 25$\mu$m/60$\mu$m colours of the entire galaxies.
%
In order to understand the Seyfert-2 dichotomy we try to
disentangle the AGN and host contributions to the IR
spectral energy distributions (SEDs) and to
explore the origin of the
cool IRAS 25$\mu$m/60$\mu$m colours. Therefore we 
consider the nuclear MIR flux concentration
(Fig.\,\ref{fig_mir_concentration}). 

Firstly, we consider the MIR concentration and the F25/F60 ratios of
the entire galaxies irrespective of the Seyfert types.
Note that AGN heated dust contributes mainly to the 3--40$\mu$m emission and
its SED decreases longward of 40$\mu$m.
The main feature of this diagram (Fig.\,\ref{fig_mir_concentration})
is 1) that sources with warm F25/F60 also have high MIR concentration
($>$20\%) and 2) that sources with low MIR concentration have cool
F25/F60, hence they lie in the lower left
corner of Figure\,\ref{fig_mir_concentration}.
These two populations can be understood in a simple scheme:
1) a powerful AGN dominates the MIR emission of the host galaxy and
leads to warm F25/F60, if it can also heat a substantial
amount of dust in the host galaxy.   
2) a nucleus, which is faint relative to the cool host,
has both a low MIR concentration {\it and}
a cool F25/F60 ratio.

\begin{figure}
  \vspace{-3mm}

  \hspace{-6mm}
  \includegraphics[angle=0,width=9.5cm,clip=true]{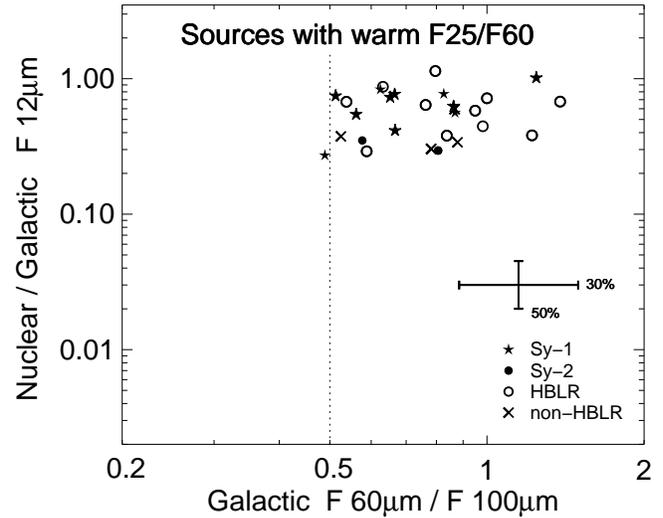}
  
  \hspace{-6mm}
  \includegraphics[angle=0,width=9.5cm,clip=true]{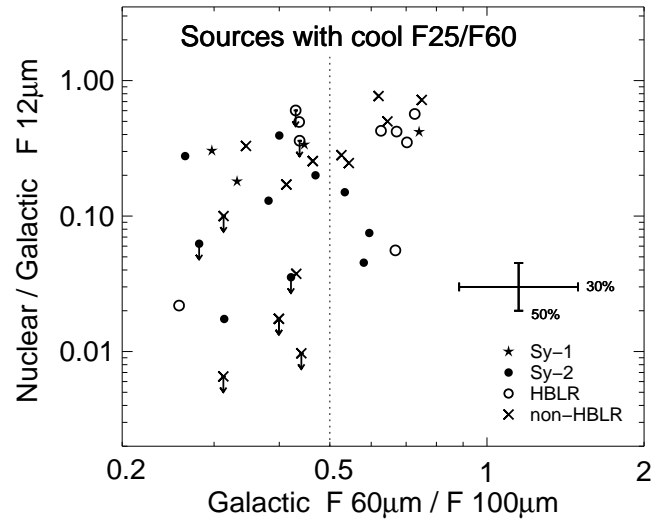}

  \caption{MIR concentration and FIR colours
    for warm (top) and cool sources (bottom).
    Symbols with arrows denote 3-$\sigma$ upper limits.
    For illustrative purpose,
    the dotted vertical line at F60/F100 = 0.5 marks
    the low end of the warm F25/F60 sources. 
    \label{fig_60_100}
  }
\end{figure}

However, the sources at {\it high} concentration {\it and}
cool F25/F60 ratio, in the upper left corner of that diagram,
are unexpected at a first glance and deserve a refined consideration.
Their nuclear contribution to the entire
F12 is relatively strong indicating a high AGN/host contrast, but 
their host SED is obviously dominated by a large amount of cold dust. 
In fact, the sources with cool F25/F60 and high nuclear MIR concentration
show also cooler F60/F100 than the warm F25/F60 sources
(Fig.\,\ref{fig_60_100}). 
Consequently the sources with
high MIR concentration ($>$20\%) may show a range of 
overall warm to cool F25/F60 and F60/F100, depending on the amount and
distribution of cold dust in the host.
Since dust mass increases with decreasing dust
temperature, the cool sources have more dust per AGN
strength than warm ones.  

Now we consider the distribution of the Seyfert types in
Figure\,\ref{fig_mir_concentration}: 
All Sy1s and virtually all (18/20) HBLR Sy2s 
emit more than 20\% of their total MIR flux in the nucleus
The non-HBLR Sy2s tend to have both cooler F25/F60 
and less concentration of MIR flux, 
reaching down to about 1\%.
The Sy2s not yet observed by spectropolarimetry are spread over the
entire range of HBLRs and non-HBLRs. 
While it is known that warm Seyferts exhibit a higher nuclear flux
concentration than cool Seyferts (e.g. Roche et al. 1991), our
diagram also shows the overall difference in the
distribution of HBLRs and non-HBLRs. 
However, if we consider the MIR concentration of
the cool sources only, then HBLRs and non-HBLRs
show more similar distributions, at least with regard to the low
number statistics. 
Notably, our sample contains also cool Sy1s, but all four of them
have high MIR concentration ($>$20\%). 
Table\,\ref{table_sample} lists the statistics of the
distribution of MIR concentration for each type. 
Many non-HBLRs have a  higher FIR luminosity and lower nuclear MIR luminosity 
than the bulk of the other Seyfert sources
(Figure\,\ref{fig_luminosities}, Table\,\ref{table_sample}).

\begin{figure}
  \vspace{-3mm}
  
  \hspace{-6mm}
  \includegraphics[angle=0,width=9.5cm,clip=true]{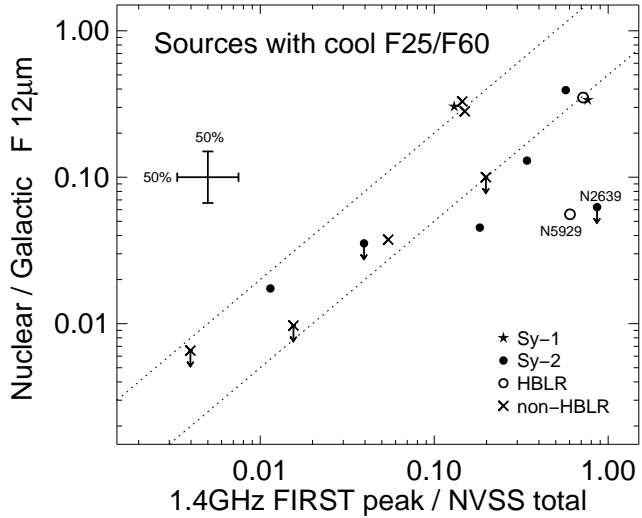}

  \caption{MIR versus radio flux concentration of cool sources (F25/F60\,$<$\,0.25).
    Symbols with arrows denote 3-$\sigma$ upper limits.
    The dotted lines mark a factor 2 around unity.
    We note that in a plot of the warm sources
    (F25/F60\,$>$\,0.25) most of them
    lie in the upper right corner within the range of the dotted lines.  
    \label{fig_radio_concentration}
  }
\end{figure}

The results remain unchanged, when 
using IRAS F12\,/\,F60 instead of IRAS F25\,/\,F60
or nuclear\,F12\,/\,galactic\,F25 instead of
nuclear\,F12\,/\,galactic\,F12.
Also we did not find any trend of MIR concentration with distance,
indicating that on the scale of 100-500 pc for nucleus and 10-30 kpc
for the entire galaxies the differences of MIR concentration are real and not
caused by an observational bias.
The same applies for the radio concentration discussed further below.

When combining the type distribution in
Figure\,\ref{fig_mir_concentration} with the SED 
properties the main conclusion is that compared with the bulk of
HBLRs and Sy1s the non-HBLRs are 
preferentially cool F25/F60 sources, where the AGN/host contrast is low or 
where the host has a higher dust mass per AGN strength. Both cases
argue in favour of observational biases as explanation for
the decreasing detection rate of polarised BLRs in cool F25/F60 sources. 

\subsection{Intrinsically weak or obscured nuclei }
\label{section_extinction}

So far, we have considered the observed properties of the sources, but
for the Seyfert-2 dichotomy intrinsic properties are relevant.
The fact that cool sources seem to have more dust than the warm ones
calls for additional clues with respect to extinction, not only of the
nucleus, but also of the more extended scattering region required for
spectropolarimetric detection of broad lines.

Moderate extinction (A$_{\rm V}$ $<$ 10) may be indicated
by the Balmer ratios of the narrow
lines, if the line fluxes are corrected for stellar absorption. But
this correction is not available for our entire sample. 
Nevertheless, the distribution of the observed Balmer ratios, as listed in
Table\,\ref{table_sources}, is indistinguishable for  
non-HBLR and HBLR Sy2s of our sample.  
This confirms 
the finding by several authors -- also based on X-ray spectra --
that non-HBLR and HBLR samples exhibit similar
column densities and obscuration (e.g. Alexander 2001, 
Gu et al. 2001, Tran 2003). 

The sources with cool IRAS colours {\it and} low
MIR concentration could contain either an {\it intrinsically} weak
AGN compared with
the host galaxy, and/or these nuclei, in principle, could
suffer from {\it extreme} MIR flux obscuration by a factor up to 1000.
In order to discriminate between these two possibilities, the 
silicate 9.7$\mu$m absorption feature may provide valuable clues.
But ground-based 8-13\,$\mu$m spectra of high-spatial resolution do not cover
a sufficient wavelength range, in order to determine the
continuum baseline free from PAH contributions (e.g. Roche et
al. 1991) so that other L or M band photometry has to be involved;
and suitable Spitzer
spectra for our sample are still being observed and under evaluation
by the proposers (e.g. Buchanan et al. 2006). 

Therefore we here consider 
-- as far as data are available --
the nuclear to total radio flux at 1.4 GHz, which arises from AGN as
well as star formation in the entire galaxies and which is essentially 
unaffected by extinction.
The spatial resolutions of the FIRST peak flux and the NVSS total flux 
are 5$\arcsec$ and 40$\arcsec$ (FWHM),
roughly comparable to the resolution of the ground-based and
IRAS MIR data, respectively. 
Figure\,\ref{fig_radio_concentration} depicts
the MIR and radio concentration 
of the cool sources. 

The sources with lowest radio concentration ($\simless$\,0.1) show also
lowest MIR concentration. 
If they were intrinsically much more luminous and
their low MIR concentration were due to MIR extinction, then -- due to
the lower radio extinction -- one
would expect to find them still at high radio concentration. 
Since this is not the case we conclude that
their AGN are {\it intrinsically} weak compared 
to the host galaxy.


But at intermediate and high radio concentration (FIRST/NVSS $\simgreat$ 0.1) 
two of the cool sources have a lower MIR concentration.
NGC2639 has a high H$_\alpha$\,/\,H$_{\beta}$ ratio ($>$\,15),
and both NGC2639 and NGC5929 show tentative 9.7\,$\mu$m silicate absorption
in their Spitzer spectra, which we inspected from the Spitzer archive.
This suggests that in these few sources a relatively
weak AGN is not the only explanation and that also extinction plays a
role.

We note in addition that most of the warm sources (F25/F60\,$>$\,0.25)
show high concentration in both MIR and radio.
If plotted in Figure\,\ref{fig_radio_concentration}, they would lie
close to unity, a few  (both HBLRs and non-HBLRs)
tending towards lower MIR than radio concentration similar as
NGC\,2639 does. 
The relation between MIR and radio concentration and the fact that
only few sources deviate from it 
also argue against the {\it general} explanation of the cool F25/F60
ratio by a high
inclination of the torus
as was recently again proposed by Zhang \& Wang (2006).

Compared with HBLRs most of the non-HBLRs
have on average about a factor 3 
lower [OIII] luminosity (Fig.\,\ref{fig_luminosities},
Table\,\ref{table_sample}).
Since this is not caused by extinction only, 
as concluded from the correlation between MIR and radio concentration
(Fig.\,\ref{fig_radio_concentration}),
they house an intrinsically weaker AGN. 

%
%

\section{Conclusions}

The science verification observations of VISIR at the VLT establish
the excellent photometric agreement with previous measurements of Seyfert
nuclei. The VISIR data suggest that almost
all of the nuclear 12\,$\mu$m flux seen in larger $\sim$5$\arcsec$ apertures
comes from a much more compact area FWHM\,$\sim$\,0$\farcs$35. 

In order to explore the nuclear dust emission of Seyfert nuclei,
we present here a comparison between mid-infrared
photometry and [OIII] at matching spatial resolution. 
The MIR\,/\,[OIII] distributions argue against the existence
of naked Seyferts, in particular when compared with
the dust-poor nucleus of the narrow-line radio galaxy M\,87.
The distributions of non-HBLRs are comparable to those
of broad- and narrow-line Sy1s and HBLRs. 
Our results suggests that all Seyfert nuclei possess the same
physical structure, where the central engine is surrounded by a
dust torus as proposed in orientation-based unified models.

While the presence of a dust torus is a necessary requirement for a hidden
BLR, our data do not allow to infer directly that such a BLR exist.
Some non-HBLRs may be misaligned narrow line
Sy1s.  On the other hand,
the cool non-HBLRs house on average an intrinsically less luminous AGN 
and show a lower AGN/host contrast. 
In addition to these observational handicaps, the non-HBLRs 
are surrounded by the same or potentially higher absolute amount of
obscuring material as do the brighter HBLR nuclei. 
This suggests that in most cases the failure to detect a hidden BLR 
in current spectropolarimetric observations is an 
apparent effect caused by observational biases.

The nuclear continuum of the non-HBLRs should be polarized, 
since the [OIII] emission has to be excited by nuclear photons. So far the 
observed polarisation is less than 0.5\%. Then the 
anticipated high nuclear polarisation may be diluted by starlight or
by interstellar polarisation in the host galaxy or by unfavourable
scattering geometries, which also prevent the detection of polarised
broad lines.

\begin{acknowledgements}
  We thank Roberto Maiolino for sending us his compilation of
  ground-based N-band observations,
  Stuart Lumsden for his critical, constructive referee report, 
  and
  Nicola Bennert, Robert Antonucci and Andreas Efstathiou for
  intriguing comments on the manuscript.
  This work substantially benefitted from SIMBAD and the NASA Extragalactic
  Database NED. 
  M.\,H. was supported  by Nordrhein--Westf\"alische Akademie der
  Wissenschaften, and H.\,H. by Deutsche Forschungsgemeinschaft via 
  Sonderforschungsbereich 439. 
\end{acknowledgements}



\begin{table*}
\normalsize
\caption{Source parameters.
}  
\label{table_sources}      
\scriptsize
\vspace{-3mm}

\begin{center}
\begin{tabular}{@{\hspace{1mm}}lccc|rc|@{\hspace{1mm}}r@{\hspace{1mm}}r@{\hspace{1mm}}r@{\hspace{1mm}}r@{\hspace{1mm}}|@{\hspace{1mm}}r@{\hspace{1mm}}r@{\hspace{1mm}}r@{\hspace{1mm}}r|@{\hspace{1mm}}r@{\hspace{1mm}}r@{\hspace{1mm}}}
(1)    &(2)& (3)  & (4)  & (5)    & (6) .& (7)     & (8)     & (9)     & (10)     & (11)   & (12)      & (13)       & (14)& (15)  & (16) \\
object & D$_{\rm L}$ & type & HBLR & F$_{\rm 10 - 12}^{\rm ~~nuc}$ & ref & IRAS 12 & IRAS 25 & IRAS 60 & IRAS 100 & [OIII] & H$_\beta$ & H$_\alpha$ & ref & FIRST & NVSS \\
       & Mpc  &      &      & mJy    &      & Jy      & Jy      & Jy      & Jy       &        & narrow    & narrow     &     & mJy   & mJy  \\
\hline
VISIR at 11.25\,$\mu$m: \\
\hline
                                    CenA &   7.8 & 2.0 &                              ? &     946 &     1 & 13.260 & 17.260 & 162.200 & 313.800 &    120 &     49 &    275 &  10 &   5700.0 & 278000.0 \\
                             ESO141-G055 & 158.4 & 1.0 &                                &     169 &     1 &  0.310 &  0.380 &   0.610 &   1.000 &   1644 &        &        &   3 &          &      6.9 \\
                                  IC5063 &  49.0 & 2.0 &                            yes &     752 &     1 &  1.110 &  3.940 &   5.870 &   4.250 &   5642 &    629 &   3711 &   3 &          &   2100.0 \\
                             MCG-3-34-64 &  71.8 & 1.8 &                            yes &     674 &     1 &  0.940 &  2.970 &   6.200 &   6.200 &  15073 &    780 &   6326 &   3 &          &    274.9 \\
                             MCG-6-30-15 &  33.4 & 2.0 &                              ? &     392 &     1 &  0.380 &  0.810 &   1.090 &   1.090 &     30 &      1 &     31 &   2 &          &      1.7 \\
                                  Mrk509 & 151.2 & 1.2 &                                &     235 &     1 &  0.315 &  0.702 &   1.364 &   1.521 &   5400 &        &        &   3 &          &     18.6 \\
          Mrk590\parbox{0cm}{$^{\rm c}$} & 115.3 & 1.2 &                                &      90 &     1 &  0.191 &  0.221 &   0.489 &   1.450 &    530 &        &        &   2 &      6.9 &     16.2 \\
                                 Mrk1239 &  86.6 & 1.5 &                           NLS1 &     660 &     1 &  0.650 &  1.140 &   1.330 &   1.070 &   4675 &        &        &   3 &     58.7 &     62.2 \\
                                 NGC2992 &  33.2 & 1.9 &                            yes &     312 &     1 &  0.630 &  1.380 &   7.510 &  17.200 &   3600 &    330 &   2299 &   2 &          &    226.2 \\
                                 NGC3783 &  42.0 & 1.0 &                                &     645 &     1 &  0.840 &  2.096 &   3.260 &   4.900 &   7626 &        &        &   3 &          &     43.6 \\
                                 NGC4507 &  51.0 & 2.0 &                            yes &     589 &     1 &  0.517 &  1.387 &   4.300 &   5.400 &   8284 &   1049 &   3947 &   3 &          &     66.1 \\
                                 NGC5427 &  37.7 & 1.9 &                              ? & $<$  10 &     1 &  0.283 &  0.623 &   6.870 &  16.310 &    540 &     54 &    160 &  12 &      3.3 &     82.8 \\
                                 NGC5995 & 110.0 & 2.0 &                            yes &     332 &     1 &  0.390 &  0.859 &   3.646 &   6.632 &    740 &    120 &   2339 &  18 &          &     30.0 \\
                                 NGC7213 &  25.1 & 1.5 &                                &     250 &     1 &  0.606 &  0.742 &   2.000 &   3.000 &   1301 &        &        &   3 &          &    247.0 \\
                                 NGC7314 &  20.5 & 1.9 &                              ? &      74 &     1 &  0.267 &  0.578 &   3.736 &  14.150 &    609 &     64 &   1300 &  12 &          &     31.0 \\
                              PG2130+099 & 282.6 & 1.0 &                                &     160 &     1 &  0.186 &  0.380 &   0.479 &   0.485 &   1042 &        &        &   3 &          &      6.0 \\
\hline
 other 10-12 $\mu$m obs. \\
\hline
                                   3C120 & 144.9 & 1.0 &                                &     109 &     4 &  0.330 &  0.710 &   1.310 &   2.640 &   3045 &        &        &   3 &        & 3439.0 \\
                                Circinus &   6.2 & 2.0 &                            yes &    4100 &     3 & 18.800 & 68.440 & 248.700 & 315.850 &    832 &     56 &   1088 &   7 &        & 1500.0 \\
                          IRAS00521-7054 & 310.4 & 2.0 &                            yes &     219 &     8 &  0.281 &  0.803 &   1.020 &   0.830 &    768 &     96 &    581 &   3 &        &   54.1 \\
                          IRAS01475-0740 &  76.7 & 2.0 &                            yes &     186 &     5 &  0.308 &$<$0.900&   1.048 &   6.500 &    530 &    101 &    777 &   3 &        &  318.2 \\
  IRAS01527+0622\parbox{0cm}{$^{\rm c}$} &  75.6 & 1.9 &                              ? &      15 &     5 &  0.200 &  0.240 &   1.190 &   2.000 &    599 &    129 &        &  12 &        &   16.0 \\
                          IRAS03362-1641 & 162.5 & 2.0 &                             no &     148 &     4 &  0.205 &  0.497 &   1.058 &   1.100 &    179 &     29 &    171 &   3 &        &    9.0 \\
                          IRAS03450+0055 & 135.9 & 1.5 &                           NLS1 &     170 &     5 &  0.283 &  0.506 &   0.470 &   0.700 &    999 &        &        &  16 &        &   32.0 \\
                          IRAS05189-2524 & 188.2 & 2.0 &                            yes &     498 &     5 &  0.740 &  3.470 &  13.250 &  12.520 &    390 &     49 &    299 &   6 &        &   28.8 \\
                          IRAS22017+0319 & 273.8 & 2.0 &                            yes &     137 &     4 &  0.287 &  0.722 &   1.160 &   1.120 &   2183 &    235 &    893 &   3 &        &   16.1 \\
  IRAS22377+0747\parbox{0cm}{$^{\rm c}$} & 109.0 & 1.8 &                              ? &      63 &     5 &  0.203 &  0.343 &   0.826 &   2.480 &   1439 &    530 &   3281 &   3 &        &   15.1 \\
                              MCG-2-8-39 & 130.9 & 2.0 &                            yes &     114 &     5 &  0.200 &  0.475 &   0.508 &   0.770 &   1829 &    117 &    365 &   3 &        &    8.7 \\
                                    Mrk3 &  58.5 & 2.0 &                            yes &     290 &     7 &  0.760 &  3.180 &   4.270 &   3.500 &  10699 &    721 &   4246 &   3 &        & 1100.4 \\
                                    Mrk9 & 176.0 & 1.5 &                                &     235 &     5 &  0.217 &  0.500 &   0.960 &   1.100 &   1089 &        &        &  12 &    2.3 &    3.6 \\
                                   Mrk79 &  96.7 & 1.2 &                                &     255 &     5 &  0.306 &  0.762 &   1.500 &   2.400 &   3700 &        &        &  12 &   10.0 &   20.5 \\
          Mrk279\parbox{0cm}{$^{\rm c}$} & 133.4 & 1.5 &                                &      76 &     7 &$<$0.205&$<$0.333&   1.255 &   2.200 &    999 &        &        &   2 &        &   23.2 \\
          Mrk334\parbox{0cm}{$^{\rm c}$} &  95.6 & 1.8 &                             no &     162 &     5 &  0.225 &  1.050 &   6.000 &   8.000 &    490 &    330 &   1599 &   2 &        &   27.9 \\
          Mrk335\parbox{0cm}{$^{\rm c}$} & 112.6 & 1.0 &                           NLS1 &     210 &     5 &  0.270 &  0.450 &   0.350 &   0.570 &   9499 &        &        &  12 &        &    7.3 \\
                                  Mrk348 &  65.1 & 2.0 &                            yes &     117 &     4 &  0.308 &  0.835 &   1.290 &   1.540 &   3594 &    258 &   1243 &   3 &        &  292.2 \\
                                 Mrk463E & 223.9 & 2.0 &                            yes &     395 &     4 &  0.500 &  1.570 &   2.184 &   1.924 &   5631 &    624 &   2643 &   2 &  349.5 &  380.5 \\
          Mrk530\parbox{0cm}{$^{\rm c}$} & 129.3 & 1.5 &                                &      77 &     7 &  0.158 &$<$0.241&   0.852 &   2.040 &    290 &        &        &   2 &        &        \\
          Mrk573\parbox{0cm}{$^{\rm c}$} &  74.6 & 2.0 &                             no &     167 &     9 &  0.280 &  0.850 &   1.240 &   1.430 &   9273 &    811 &   3166 &   3 &        &   24.0 \\
                                  Mrk607 &  38.3 & 2.0 &                             no &     100\parbox{0cm}{$^{\rm s}$} &     5 &  0.330 &  1.060 &   2.150 &   2.750 &   1224 &    135 &    576 &   3 &        &    6.0 \\
                                  Mrk618 & 156.4 & 1.0 &                                &     270 &     5 &  0.336 &  0.788 &   2.700 &   4.240 &   1602 &        &        &   3 &        &   17.0 \\
                                  Mrk704 & 128.0 & 1.5 &                                &     270 &     5 &  0.350 &  0.530 &   0.400 &   0.360 &    850 &        &        &   2 &    4.9 &    6.1 \\
          Mrk789\parbox{0cm}{$^{\rm c}$} & 138.0 & 1.0 &                                &      55 &     5 &  0.145 &  0.617 &   3.300 &   5.000 &    450 &        &        &   2 &   25.4 &   35.2 \\
          Mrk817\parbox{0cm}{$^{\rm c}$} & 138.0 & 1.5 &                                &     220 &     5 &  0.380 &  1.420 &   2.330 &   2.350 &   1400 &        &        &   2 &    8.1 &   11.2 \\
          Mrk841\parbox{0cm}{$^{\rm c}$} & 160.3 & 1.5 &                                &     139 &     5 &  0.192 &  0.431 &   0.600 &   0.480 &   2499 &        &        &   2 &        &  -14.8 \\
                                  Mrk897 & 115.1 & 2.0 &                              ? &      60 &     5 &  0.240 &  0.500 &   2.800 &   4.670 &    877 &     58 &    177 &   5 &        &   16.9 \\
          Mrk993\parbox{0cm}{$^{\rm c,d}$} &  67.3 & 2.0 &                              ? &      18 &     9 &$<$0.131&$<$0.129&   0.296 &   1.320 &    299 &     51 &    279 &   2 &        &    5.4 \\
                                 Mrk1040 &  72.2 & 1.5 &                                &     333 &     5 &  0.610 &  1.315 &   2.555 &   4.551 &    748 &        &        &   3 &        &   13.3 \\
                                  NGC424 &  50.9 & 2.0 &                            yes &     490 &    12 &  1.100 &  1.740 &   1.790 &   1.823 &   4199 &    910 &        &  12 &        &   23.0 \\
                                  NGC513 &  85.0 & 2.0 &                            yes & $<$ 100 &     4 &  0.166 &  0.277 &   1.935 &   4.500 &    350 &     51 &    264 &  18 &        &   52.9 \\
                                  NGC985 & 190.8 & 1.0 &                                &     143 &     5 &  0.207 &  0.523 &   1.381 &   1.890 &   2220 &        &        &   3 &    4.7 &   15.3 \\
         NGC1068\parbox{0cm}{$^{\rm c}$} &  16.3 & 2.0 &                            yes &   25500 &     4 & 39.840 & 87.570 & 196.370 & 257.370 &  48336 &   3039 &  13041 &   3 & 1855.3 & 4848.1 \\
                                 NGC1241 &  58.5 & 2.0 &                             no &      79 &     5 &  0.240 &  0.447 &   3.557 &  10.300 &   3700 &    660 &   1920 &  11 &    2.9 &   20.0 \\
                                 NGC1365 &  23.5 & 1.8 &                              ? &     510 &     2 &  3.400 & 10.080 &  76.100 & 142.500 &    619 &    338 &   2960 &  10 &        &  375.9 \\
                                 NGC1386 &  12.4 & 2.0 &                             no &     400 &     3 &  0.520 &  1.460 &   5.920 &   9.550 &   7999 &    530 &   2489 &   1 &        &   37.1 \\
                                 NGC1667 &  65.7 & 2.0 &                             no & $<$   8 &     5 &  0.430 &  0.677 &   5.950 &  14.900 &    637 &    183 &   1785 &   9 &        &   75.8 \\
                                 NGC2110 &  33.6 & 2.0 &                            yes &     198 &    11 &  0.349 &  0.840 &   4.130 &   5.680 &   1700 &    270 &   2200 &   2 &        &  298.8 \\
                                 NGC2273 &  26.4 & 2.0 &                            yes &     185 &    10 &  0.440 &  1.360 &   6.410 &   9.550 &   3299 &    330 &        &  12 &        &   62.6 \\
                                 NGC2639 &  48.1 & 1.9 &                              ? & $<$  10 &     5 &  0.160 &  0.209 &   1.980 &   7.050 &    139 &     11 &    179 &   4 &   99.3 &  115.0 \\
                                 NGC3031 &   3.3 & 1.8 &                            yes &     128 &     4 &  5.860 &  5.420 &  44.730 & 174.000 &    999 &    306 &   2800 &   4 &        &   86.0 \\
         NGC3080\parbox{0cm}{$^{\rm c}$} & 155.8 & 1.0 &                                &      16 &     5 &$<$0.091&$<$0.153&   0.349 &   0.874 &    129 &        &        &   2 &        &    2.9 \\
                                 NGC3081 &  34.4 & 2.0 &                            yes &     436 &    12 &  0.500 &  0.940 &   3.160 &$<$5.000 &   3899 &    290 &   1300 &   2 &        &    5.4 \\
                                 NGC3185 &  17.5 & 2.0 &                              ? &      20 &     5 &  0.154 &  0.140 &   1.400 &   3.670 &    540 &     94 &    637 &   4 &    1.8 &    5.3 \\
         NGC3227\parbox{0cm}{$^{\rm c}$} &  16.6 & 1.5 &                                &     225 &     4 &  0.667 &  1.760 &   7.820 &  17.500 &   8199 &        &        &   2 &   74.4 &   97.5 \\
                                 NGC3281 &  46.1 & 2.0 &                             no &     309 &    12 &  0.910 &  2.630 &   6.930 &   7.890 &    550 &     49 &        &   2 &        &   80.1 \\
         NGC3362\parbox{0cm}{$^{\rm c}$} & 120.9 & 2.0 &                             no &      12 &     5 &  0.150 &  0.350 &   2.130 &   3.150 &    499 &     49 &    252 &  13 &    5.5 &   16.6 \\
         NGC3516\parbox{0cm}{$^{\rm c}$} &  38.1 & 1.5 &                                &     230 &     7 &  0.410 &  1.010 &   1.850 &   2.130 &   2700 &        &        &   2 &        &   31.3 \\
                                 NGC3660 &  53.1 & 2.0 &                             no &      33 &     5 &  0.193 &  0.224 &   1.870 &   4.530 &    335 &    111 &    559 &   9 &        &   12.3 \\
         NGC3786\parbox{0cm}{$^{\rm c}$} &  38.5 & 1.8 &                              ? &      59 &     5 &  0.150 &  0.200 &   2.000 &   5.000 &    839 &     99 &        &  12 &   10.6 &   18.6 \\
         NGC3982\parbox{0cm}{$^{\rm c}$} &  15.9 & 2.0 &                             no &      19 &     5 &  0.507 &  0.833 &   6.567 &  15.230 &   1879 &     68 &    490 &   4 &    3.1 &   56.4 \\
         NGC4051\parbox{0cm}{$^{\rm c}$} &  10.0 & 1.5 &                           NLS1 &     260 &     7 &  0.855 &  1.590 &   7.100 &  23.900 &   3899 &        &        &   4 &   12.3 &   94.4 \\
         NGC4151\parbox{0cm}{$^{\rm c}$} &  14.3 & 1.5 &                                &    1520 &     4 &  1.968 &  4.833 &   6.315 &   7.640 & 115999 &        &        &   4 &  290.8 &  359.6 \\
         NGC4235\parbox{0cm}{$^{\rm c}$} &  34.7 & 1.0 &                                &      34 &     5 &  0.125 &  0.156 &   0.316 &   0.646 &    240 &        &        &  12 &    4.7 &   12.2 \\
         NGC4253\parbox{0cm}{$^{\rm c}$} &  55.9 & 1.0 &                                &     241 &     5 &  0.385 &  1.290 &   4.020 &   4.660 &   4535 &        &        &   3 &   38.2 &   38.1 \\
         NGC4388\parbox{0cm}{$^{\rm c}$} &  36.3 & 2.0 &                            yes &     290 &     3 &  0.996 &  3.420 &  10.000 &  17.000 &   5642 &    479 &   2744 &   3 &   24.4 &  119.4 \\
         NGC4501\parbox{0cm}{$^{\rm c}$} &  32.8 & 2.0 &                             no & $<$  15 &     5 &  2.290 &  2.980 &  19.680 &  62.970 &    340 &     27 &    229 &   4 &    1.1 &  277.1 \\
                                 NGC4593 &  38.8 & 1.0 &                                &     257 &     4 &  0.344 &  0.808 &   3.050 &   5.947 &   1343 &        &        &   3 &    2.8 &    4.4 \\
                                 NGC4968 &  42.6 & 2.0 &                              ? &     115 &     5 &  0.390 &  1.000 &   2.370 &   2.940 &   1773 &    165 &    989 &   3 &        &   34.5 \\
\hline
\end{tabular}	
\end{center}
\end{table*}

\addtocounter{table}{-1}
\begin{table*}
\normalsize
\caption{continued. 
}  
\scriptsize
\vspace{-3mm}

\begin{center}
\begin{tabular}{@{\hspace{1mm}}lccc|rc|@{\hspace{1mm}}r@{\hspace{1mm}}r@{\hspace{1mm}}r@{\hspace{1mm}}r@{\hspace{1mm}}|@{\hspace{1mm}}r@{\hspace{1mm}}r@{\hspace{1mm}}r@{\hspace{1mm}}r|@{\hspace{1mm}}r@{\hspace{1mm}}r@{\hspace{1mm}}}
(1)    &(2)& (3)  & (4)  & (5)    & (6) .& (7)     & (8)     & (9)     & (10)     & (11)   & (12)      & (13)       & (14)& (15)  & (16) \\
object & D$_{\rm L}$ & type & HBLR & F$_{\rm 10 - 12}^{\rm ~~nuc}$ & ref & IRAS 12 & IRAS 25 & IRAS 60 & IRAS 100 & [OIII] & H$_\beta$ & H$_\alpha$ & ref & FIRST & NVSS \\
       & Mpc  &      &      & mJy    &      & Jy      & Jy      & Jy      & Jy       &        & narrow    & narrow     &     & mJy   & mJy  \\
\hline
         NGC5033\parbox{0cm}{$^{\rm c}$} &  12.5 & 1.8 &                              ? &      24 &     5 &  1.380 &  2.000 &  16.000 &  51.000 &    530 &     55 &    530 &   2 &    1.4 &  121.5 \\
                                 NGC5135 &  59.3 & 2.0 &                             no &     157 &     5 &  0.638 &  2.380 &  16.800 &  30.900 &   2190 &    426 &   2610 &   8 &        &  199.8 \\
         NGC5194\parbox{0cm}{$^{\rm c}$} &   6.6 & 2.5 &                             no & $<$  70 &     4 &  7.210 &  9.560 &  97.000 & 220.000 &   1199 &     80 &    700 &   4 &    6.6 &  427.1 \\
         NGC5252\parbox{0cm}{$^{\rm c}$} &  99.8 & 1.9 &                            yes &      28 &     5 &  0.080 &  0.140 &   0.850 &   1.210 &   2190 &    370 &   1380 &  10 &   11.7 &   16.3 \\
         NGC5256\parbox{0cm}{$^{\rm c}$} & 121.9 & 2.0 &                             no &      30 &     5 &  0.230 &  0.980 &   7.200 &  10.500 &    229 &     49 &    231 &   4 &        &        \\
         NGC5273\parbox{0cm}{$^{\rm c}$} &  15.3 & 1.9 &                              ? &      42 &     5 &  0.120 &  0.290 &   0.900 &   1.560 &    220 &     20 &     66 &   2 &    2.6 &    3.5 \\
         NGC5283\parbox{0cm}{$^{\rm c}$} &  44.9 & 2.0 &                             no &      15 &     5 &  0.040 &  0.130 &   0.210 &   0.400 &    870 &    108 &    345 &  14 &        &   13.2 \\
         NGC5347\parbox{0cm}{$^{\rm c}$} &  33.6 & 2.0 &                            yes &     208 &     5 &  0.308 &  0.962 &   1.420 &   2.640 &    447 &     49 &    270 &   3 &    3.2 &    5.6 \\
                                 NGC5506 &  26.6 & 1.9 &    yes\parbox{0cm}{$^{\rm x}$} &     720 &     5 &  1.240 &  4.170 &   8.420 &   8.870 &   5212 &    569 &   3920 &   9 &  310.0 &  338.8 \\
         NGC5548\parbox{0cm}{$^{\rm c}$} &  74.5 & 1.5 &                                &     292 &     4 &  0.400 &  0.769 &   1.070 &   1.640 &   3599 &        &        &   3 &   11.0 &   28.2 \\
                                 NGC5643 &  17.2 & 2.0 &                             no &     310 &     3 &  1.100 &  3.600 &  20.000 &  38.000 &   7999 &    480 &   2900 &   2 &   30.0 &  200.0 \\
         NGC5674\parbox{0cm}{$^{\rm c}$} & 108.8 & 1.9 &                              ? &      24 &     5 &  0.144 &  0.280 &   1.400 &   3.700 &   1350 &    338 &   1659 &  10 &    4.6 &   34.2 \\
         NGC5695\parbox{0cm}{$^{\rm c}$} &  61.0 & 2.0 &                             no & $<$  10 &     5 &  0.100 &  0.128 &   0.560 &   1.790 &    550 &     40 &    199 &  13 &    1.2 &    6.3 \\
         NGC5929\parbox{0cm}{$^{\rm c}$} &  35.8 & 2.0 &                            yes &      24 &     5 &  0.430 &  1.620 &   9.140 &  13.690 &    930 &    240 &   1300 &   2 &   65.4 &  108.4 \\
         NGC5940\parbox{0cm}{$^{\rm c}$} & 149.1 & 1.0 &                                &      26 &     5 &$<$0.167&  0.115 &   0.743 &   1.750 &    350 &        &        &  17 &        &    9.0 \\
                                 NGC5953 &  28.2 & 2.0 &                              ? &      24 &     5 &  0.530 &  1.160 &  11.550 &  19.890 &    629 &    330 &   2399 &  12 &   16.7 &   91.4 \\
         NGC6104\parbox{0cm}{$^{\rm c}$} & 123.0 & 1.5 &                                &      12 &     5 &$<$0.980&$<$0.840&   0.500 &   1.750 &    132 &        &        &  13 &        &    6.4 \\
                                 NGC6814 &  22.4 & 1.5 &                                &     166 &     6 &  0.920 &  1.040 &   6.530 &  19.670 &   1199 &        &        &   2 &        &   49.7 \\
                                 NGC7130 &  70.0 & 2.0 &                             no &     290 &    13 &  0.580 &  2.160 &  16.700 &  25.900 &    270 &     89 &    580 &   6 &        &  189.7 \\
                                 NGC7172 &  37.4 & 2.0 &                             no &     107 &     4 &  0.420 &  0.880 &   5.760 &  12.420 &     99 &     20 &    140 &  19 &        &   36.8 \\
                                 NGC7465 &  28.3 & 2.0 &                              ? &      52 &     5 &  0.260 &  0.480 &   3.820 &   8.140 &    383 &    137 &    660 &  15 &        &   19.1 \\
         NGC7469\parbox{0cm}{$^{\rm c}$} &  70.8 & 1.2 &                                &     565 &     2 &  1.350 &  5.789 &  25.870 &  34.900 &   8399 &        &        &   2 &        &  180.5 \\
                                 NGC7582 &  22.6 & 2.0 &    yes\parbox{0cm}{$^{\rm x}$} &     690 &     3 &  1.620 &  7.400 &  52.000 &  83.000 &   2999 &   1199 &   9999 &   2 &        &  270.0 \\
         NGC7674\parbox{0cm}{$^{\rm c}$} & 126.6 & 2.0 &                            yes &     260 &     3 &  0.680 &  1.920 &   5.360 &   8.300 &   7178 &    637 &   3306 &   3 &        &  220.9 \\
         NGC7682\parbox{0cm}{$^{\rm c}$} &  74.4 & 2.0 &                            yes & $<$  18 &     5 &  0.050 &  0.080 &   0.350 &   0.800 &   2299 &    270 &   1199 &  10 &        &   59.8 \\
         UGC6100\parbox{0cm}{$^{\rm c}$} & 129.2 & 2.0 &                             no &      34 &     5 &  0.145 &  0.202 &   0.574 &   1.500 &   2900 &    209 &    949 &  18 &    7.3 &   11.3 \\
\hline
\end{tabular}	
\end{center}
\normalsize

Col(1): sources marked by $^{\rm c}$ are contained in the CfA Seyfert
sample (Huchra \& Burg 1992).

$^{\rm d}$ dropped from the analysis because of meaningless IRAS
F12 and F25 upper limits, which are high relative to other quantities. 

Col (2): We used a $\Lambda$ cosmology with 
H$_0$ = 71 km\,s$^{-1}$\, Mpc$^{-1}$, $\Omega_{{\rm matter}}$ = 0.27
and $\Omega_{\Lambda}$ = 0.73. 

Col (4): references are  Heisler et al. (1997), Lumsden et
al. (2001, 2004), Moran et al. (2000, 2001, 2007), Tran (2003), see
compilation by Gu \& Huang (2002).
Sources marked by $^{\rm x}$ show broad Pa$\beta$ lines in   
ordinary near-infrared spectroscopy,
but no optical spectropolarimetric BLR
(NGC\,5506: Nagar et al. 2002, NGC\,7582: Lumsden priv. comm.,
see also Aretxaga et al. 1999). 

Col (5): in case of non-detections 
3-$\sigma$ upper limits are listed.
$^{\rm *}$ denotes barely resolved sources. 
$^{\rm s}$ value derived from Spitzer-IRS, since the 
28\,mJy in 6$\arcsec$ aperture by Maiolino et al. (1995) appear too low. 

Col (6): references are 
1 = VISIR, this work
2 = Galliano et al. (2005b, 12\,$\mu$m TIMMI2/ESO 3.6m), 
3 = Siebenmorgen et al. (2004, 12\,$\mu$m TIMMI2/ESO 3.6m), 
4 = Gorjian et al. (2004, 10\,$\mu$m MIRLIN/Palomar 5m),
5 = Maiolino et al. (1995, 5$\farcs$3 10\,$\mu$m Bolometer/MMT),
6 = Glass et al. (1982, 7$\farcs$5 12\,$\mu$m Bolometer/ESO 3.6m),
7 = Rieke (1978),
8 = Frogel \& Elias (1987), 
9 = Edelson et al. (1987),  
10 = Devereux  (1987),  
11 = Lawrence et al. (1985),   
12 = Boisson \& Durret (1986), 
13 = Wynn-Williams \& Becklin (1993).

Cols (7-10):   IRAS photometry was taken from NED, mostly Sanders et al. 2003, if available, otherwise from Faint Source Catalog,
  the source of photometry had no significant effect on the analysed quantities F25/F60 or nuclear/galactic F12. 
For some sources the photometry was estimated/improved:
for
NGC\,3362, NGC\,5252, NGC\,5283, Mrk\,334, IRAS\,01527+0622 from ISO 
(P\'erez Garcia \& Rodr\'iguez Espinosa 2001), 
and for Mrk\,897, NGC\,2992, NGC\,3081, NGC\,3783, NGC\,3786,
NGC\,5427, NGC\,7213, NGC\,7682 from Spitzer data (this work).

%

Col (11-13): the line fluxes are listed
in units 10$^{\rm -16}$  erg/s/cm$^{\rm 2}$.

Col (14): References are
 1 = Bennert et al. (2006), 
 2 = Dahari \& de Robertis (1988), 
 3 = de Grijp et al. (1992), 
 4 = Ho et al. (1995),
 5 = Kewley et al. (2001),
 6 = Kim  et al. (1995),
 7 = Oliva et al. (1994), 
 8 = Phillips et al. (1983), 
 9 = Storchi-Bergmann et al. (1995),  
10 = V\'eron-Cetty \& V\'eron (2006),
11 = Vaceli et al. (1997),
12 = Whittle (1992), 
13 = SDSS DR5, 
14 = SAO Z-machine archive (http://tdc-www.harvard.edu/cgi-bin/arc/zsearch), 
15 = Moustakas \& Kennicutt (2006),
16 = Boroson \& Meyers (1992),
17 = Bonatto \& Pastoriza (1997),
18 = Tran (2003),
19 = Sharples et al. (1984).
\end{table*}

\end{document}